\documentclass[aps,prb,twocolumn,showpacs,groupedaddress]{revtex4-2}
\usepackage{graphicx}
\usepackage{epstopdf} 
\usepackage{bm}
\begin{document}

\title{Dicke superradiance from a plasmonic nanocomposite slab}

% \affiliation command applies to all authors since the last
% \affiliation command. The \affiliation command should follow the
% other information
% \affiliation can be followed by \email, \homepage, \thanks as well.
\author{V.G.~Bordo}
\email{bordo@sdu.dk}

\affiliation{SDU Electrical Engineering, University of Southern Denmark, Alsion 2, DK-6400 S{\o}nderborg, Denmark}

%Collaboration name if desired (requires use of superscriptaddress
%option in \documentclass). \noaffiliation is required (may also be
%used with the \author command).
%\collaboration can be followed by \email, \homepage, \thanks as well.
%\collaboration
%\noaffiliation

\date{\today}

\begin{abstract}
The Dicke superradiance from an optically thin nanocomposite slab represented by metal nanoparticles dispersed in a dielectric matrix is predicted and its theory is developed from first principles. It is shown that the superradiance signal evolution is determined by the eigenvalues of the field susceptibility tensor for the slab. The excitation of the system by a pumping pulse in different polarizations as well as in the attenuated total reflection configuration is considered. It is demonstrated that the relaxation rates are enhanced when surface plasmon polaritons are excited at the interface between the substrate and superstrate. These findings can pave a way to the extension of the optical techniques based on the Dicke superradiance to the femtosecond time domain.
\end{abstract}
% insert suggested PACS numbers in braces on next line
% insert suggested keywords - APS authors don't need to do this
%\keywords{}
%\maketitle must follow title, authors, abstract, \pacs, and \keywords
\maketitle
% body of paper here - Use proper section commands
% References should be done using the \cite, \ref, and \label commands
\section{Introduction}
A gas of molecules interacting with each other through a common radiation field and confined to a container the dimensions of which are small compared with the radiation wavelength spontaneously emits coherent light, the corresponding emission rate being proportional to the number of excited molecules. This effect predicted by Dicke in 1954 \cite{Dicke54} received later on the name "Dicke superradiance". Its underlying mechanism is a spontaneous establishment of correlations between individual radiators which has a close analogy with synchronization and self-organization among oscillators in classical systems \cite{Cong16}. It has been observed in a plenty of atomic and molecular systems \cite{Andreev80} and obtained a profound theoretical description \cite{Eberly71,Haroche82,Zheleznyakov89}. Being initially investigated in atomic vapors and molecular gases, the phenomenon of superradiance has been subsequently extended to a variety of molecular systems and excitations in solids such as molecular centers, molecular aggregates and crystals, semiconductor quantum dots and nanocrystals, excitons in semiconductor quantum wells, cyclotron resonances and intersubband transitions in quantum wells \cite{Cong16}.\\
Superradiance is characterized by a high directionality, a quadratic dependence of the maximum intensity on the particle
density, and an inverse proportionality between the superradiant pulse length and the particle density. These features dictate the importance of this effect for applications in generation of intense coherent pulses whose intensity and length can be varied over broad ranges in a rather simple manner \cite{Andreev80}. Superradiance combined with stimulated emission in a laser cavity is a key mechanism in the operation of the so-called superradiant laser \cite{Thompson12}. Such a laser has an extreme spectral purity which is crucial for the stability of passive atomic clocks. \\
From the dynamical viewpoint superradiance is a transient process, which occurs over times shorter than the relaxation time of the atomic dipole moment. Therefore the extension of the optical techniques based on superradiance to short-living excitations in solids leads to the possibility of generating ultra-short pulses. For example, for the infrared transitions in atomic vapors and molecular gases the superradiant pulse duration is of the order of 1 ns \cite{Andreev80}, while for excitations in solids it can be of the order of 1 ps \cite{Cong16}.\\
The ultimately short relaxation times, which have the order of a few femtoseconds, one finds in metals where electronic excitations are effectively quenched by electron collisions. They can be further reduced in metal nanoparticles (MNPs) due to electron collisions with the particle boundary \cite{Kreibig95}. MNPs support collective electron excitations, which are known as localized surface plasmons (LSPs), with broad resonances in the optical spectral range. In the vicinity of the LSP resonance a MNP has much in common with a quantum emitter. In the quantum description, the light radiated by a MNP can be regarded as being originating from a transition between a single-plasmon and a vacuum plasmonic Fock states \cite{Bordo19}. Such a transition has a giant dipole moment ($\sim$ 10$^3$ Debye) which dictates a very large radiative decay rate of the order of $10^{15}$ s$^{-1}$ \cite{Bordo19A}.\\
MNPs can be dispersed in different dielectric matrices thus forming plasmonic nanocomposites which can be deposited on a substrate as a thin film with a thickness of a few tens nanometers \cite{Elbahri14,Tamulevicius18,Santos18}. In this context, one can expect that the radiation from MNPs distributed in a nanocomposite film will resemble superradiance from an ensemble of atoms confined to a subwavelength slab. Despite such a simple analogy, the Dicke superradiance from a subwavelength array of MNPs, to the best of our knowledge, has never been discussed in the literature. This effect is principally different from the so-called plasmonic Dicke effect \cite{Shahbazyan09} where 
an ensemble of quantum emitters located in the near field of {\it a single} MNP radiates cooperatively.\\
In the present paper, we develop the theory of the Dicke superradiance from an optically thin plasmonic nanocomposite slab from first principles. The theory is based on a rigorous treatment of the local field in the slab \cite{Bordo18} which inherently contains correlations between individual MNPs. The paper is organized as follows. Section \ref{sec:model} introduces the theoretical model adopted in the paper. In Sec. \ref{sec:evolution} the evolution of the MNPs polarization excited by a pump pulse is considered. In Sec. \ref{sec:eigenvalues} the eigenvalues of the field susceptibility tensor which determine the polarization evolution are investigated. In Sec. \ref{sec:radiation} the expression for the Dicke superradiance intensity is obtained which is numerically analyzed for different scenarios in Sec. \ref{sec:numerical}. Section \ref{sec:conclusion} summarizes the main results of the paper.\\
\section{Theoretical model}\label{sec:model}
\begin{figure}
\includegraphics[width=\linewidth]{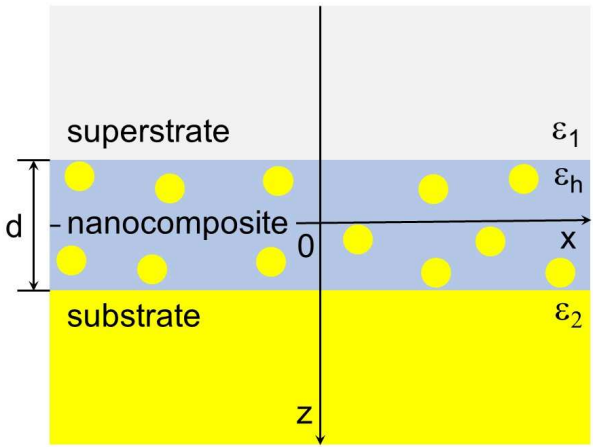}
\caption{\label{fig:scheme} The structure under consideration.}
\end{figure}

Let us consider a nanocomposite slab placed between a superstrate with the dielectric function $\epsilon_1$ and a substrate with the dielectric function $\epsilon_2$ (Fig. \ref{fig:scheme}). We assume that the slab is represented by a dielectric host medium with the dielectric function $\epsilon_h$ in which spherical MNPs of radius $R$ are randomly distributed with the number density $N$ or, equivalently, with the volume fraction $f=(4\pi/3)R^3N$. The slab is assumed to be optically thin, i.e. its thickness, $d$, is much less than the wavelength of the incident light, $\lambda$.  We direct the $z$ coordinate axis from the superstrate to the substrate along the normal to the slab boundaries and place its origin at the midpoint between them.\\
In the vicinity of the resonance with the MNP localized surface plasmon frequency, $\omega_0$, the evolution of the MNP dipole moment, ${\bf p}({\bf r},t)$, can be described in the damped harmonic oscillator model as follows
\begin{equation}\label{eq:oscil}
\frac{d^2{\bf p}({\bf r},t)}{dt^2}+\Gamma\frac{d{\bf p}({\bf r},t)}{dt}+\omega_0^2{\bf p}({\bf r},t)=a{\bf E}^{\prime}({\bf r},t),
\end{equation}
where $\Gamma$ is the damping constant and the parameter $a$ specifies the coupling between the MNP and the microscopic (local) field in the slab, ${\bf E}^{\prime}$. A comparison with the quasi-static limit \cite{Bordo16} reveals that the quantity $\Gamma$ in Eq. (\ref{eq:oscil}) coincides with the relaxation constant in the Drude model for the MNP dielectric function
\begin{equation}\label{eq:drude}
\epsilon(\omega)=\epsilon_{\infty}-\frac{\omega_p^2}{\omega(\omega+i\Gamma)},
\end{equation}
where $\epsilon_{\infty}$ is the offset originating from the interband transitions and $\omega_p$ is the metal plasma frequency, while the LSP frequency and the coupling constant are given by
\begin{equation}
\omega_0=\frac{\omega_p}{\sqrt{\epsilon_{\infty}+2\epsilon_h}}
\end{equation}
and
\begin{equation}
a=\frac{3}{4\pi}f\epsilon_h\omega_0^2,
\end{equation}
respectively.\\
The local field in Eq. (\ref{eq:oscil}) in turn can be written as a sum of the incident field in the slab, ${\bf E}({\bf r},t)$, and the field scattered by the induced dipoles as follows \cite{Born70,Bordo18}
\begin{equation}\label{eq:local}
{\bf E}^{\prime}({\bf r},t)={\bf E}({\bf r},t)+\int_{V^{\prime}} \bar{\bf F}({\bf r},{\bf r}^{\prime};\omega){\bf P}({\bf r}^{\prime},t)d{\bf r}^{\prime},
\end{equation}
where ${\bf P}({\bf r},t)=N{\bf p}({\bf r},t)$ is the polarization of the MNP dipoles. Here $\bar{\bf F}({\bf r},{\bf r}^{\prime};\omega)$ is the so-called field susceptibility tensor \cite{Sipe84,Nha96} which relates the electric field at the point ${\bf r}$ generated by a classical dipole, oscillating at frequency $\omega$, with the dipole moment itself, located at ${\bf r}^{\prime}$, $V^{\prime}$ denotes the slab volume after removal of a small volume around the MNP under consideration that excludes the singularity associated with the self-action of MNPs. This quantity can be expressed in terms of the Green's function of the vector wave equation with appropriate boundary conditions \cite{Bordo16}. When writing Eq. (\ref{eq:local}) one formally treats the quantity ${\bf P}({\bf r},t)$ as a continuous function of ${\bf r}$ that is a good approximation when the average distance between MNPs is much less than $\lambda$. The field susceptibility tensor can be represented as the Fourier integral 
\begin{eqnarray}\label{eq:fourier}
\bar{\bf F}({\bf r},{\bf r}^{\prime};\omega)\nonumber\\
=\frac{1}{(2\pi)^2}\int\bar{\bf f}(z,z^{\prime};\kappa _x,\kappa _y;\omega)
e^{i\kappa _x(x-x^{\prime})}e^{i\kappa _y(y-y^{\prime})}d\kappa _xd\kappa _y,\nonumber\\
\end{eqnarray}
where $\bar{\bf f}(z,z^{\prime};\kappa _x,\kappa _y;\omega)$ is the Fourier transform. This representation allows one to write the dipole field as a superposition of plane waves.\\
\section{Evolution of the nanoparticles polarization}\label{sec:evolution}
We assume that the incident field transmitted into the slab oscillates at the frequency $\omega$ which is close to $\omega_0$ and propagates along the $x$-axis so that
\begin{equation}\label{eq:incident}
{\bf E}({\bf r},t)={\bf E}_0(t)\exp(ik_0x-i\omega t),
\end{equation}
where ${\bf E}_0(t)$ is the amplitude envelope function and $k_0$ is the wave vector. Accordingly, we seek the solutions for the polarization and the local field in the forms
\begin{equation}\label{eq:polarization}
{\bf P}({\bf r},t)={\bf P}_0(z,t)\exp(ik_0x-i\omega t)
\end{equation}
and 
\begin{equation}\label{eq:local_0}
{\bf E}^{\prime}({\bf r},t)={\bf E}^{\prime}_0(z,t)\exp(ik_0x-i\omega t),
\end{equation}
respectively, where ${\bf P}_0(z,t)$ and ${\bf E}^{\prime}_0(z,t)$ are the time-dependent amplitudes which vary much slower than $\exp(-i\omega t)$ and we have taken into account the translational invariance of the problem along the $y$ axis. \\
Substituting Eqs. (\ref{eq:incident}) - (\ref{eq:local_0}) in Eqs. (\ref{eq:oscil}) and (\ref{eq:local}) and neglecting the first and second time derivatives of ${\bf P}_0(z,t)$ in comparison with $\omega{\bf P}_0(z,t)$ and $\omega^2{\bf P}_0(z,t)$, respectively, one comes to the equations
\begin{equation}\label{eq:beta}
\frac{d{\bf P}_0(z,t)}{dt}+\left(\frac{\Gamma}{2}-i\Delta\right){\bf P}_0(z,t)=i\beta{\bf E}^{\prime}_0(z,t),
\end{equation}
with $\beta=a/(2\omega_0)=(3/8\pi)f\epsilon_h\omega_0$, $\Delta=\omega-\omega_0$ being the resonance detuning and
\begin{equation}\label{eq:integral}
{\bf E}_0^{\prime}(z,t)={\bf E}_0(t)+\int_{-d/2}^{d/2} \bar{\bf f}(z,z^{\prime};k_0,0;\omega_0){\bf P}_0(z^{\prime},t)dz^{\prime},
\end{equation}
where we have used that $\omega\approx\omega_0$.\\
Taking into account that the nanocomposite slab under consideration has the thickness much less that the wavelength, one can neglect the $z$-dependence in the above equations and approximate the kernel in the integral equation (\ref{eq:integral}) as follows
\begin{equation}
\bar{\bf f}(z,z^{\prime};k_0,0;\omega_0)\approx \bar{\bf f}(0,0;k_0,0;\omega_0)\equiv \bar{\bf f}_0(k_0,\omega_0).
\end{equation}
This reduces Eqs. (\ref{eq:beta}) and (\ref{eq:integral}) to a single differential equation
\begin{equation}\label{eq:differ}
\frac{d{\bf P}_0(t)}{dt}+\bar{\sigma}{\bf P}_0(t)=i\beta{\bf E}_0(t),
\end{equation}
where the matrix elements of the tensor $\bar{\sigma}$ are defined as ($i$ and $j$ run the values $x,y$ and $z$)
\begin{equation}
\sigma_{ij}=\left(\frac{\Gamma}{2}-i\Delta\right)\delta_{ij}-id\beta f_{0,ij}(k_0,\omega_0).
\end{equation}
Equation (\ref{eq:differ}) can be conveniently solved by means of the Laplace transform
\begin{equation}
{\bf \Pi}(s)=\int_0^{\infty}{\bf P}_0(t)\exp(-st)dt.
\end{equation}
Due to the block diagonal structure of the matrix $\sigma_{ij}$ the problem is split into two separate problems for the incident field polarized in the plane of incidence (TM polarization) and perpendicular to it (TE polarization). The corresponding solutions are given by
\begin{equation}\label{eq:Pix}
\Pi^{TM}_x(s)=\frac{i\beta \Sigma_x(s)E_0(s)}{[s+(\Gamma _-/2)-i\Delta _-][s+(\Gamma _+/2)-i\Delta _+]},
\end{equation}
\begin{equation}\label{eq:Piz}
\Pi^{TM}_z(s)=\frac{i\beta \Sigma_z(s)E_0(s)}{[s+(\Gamma _-/2)-i\Delta _-][s+(\Gamma _+/2)-i\Delta _+]}
\end{equation}
and
\begin{equation}\label{eq:Piy}
\Pi^{TE}_y(s)=\frac{i\beta E_0(s)}{s+(\Gamma^{\prime}/2)-i\Delta^{\prime}},
\end{equation}
where we have assumed a zero polarization at $t=0$. Here $E_0(s)$ is the Laplace transform of the amplitude $E_0(t)$ and the following notations are introduced:
\begin{eqnarray}
\Sigma_x(s)=(s+\sigma _{zz})\cos\theta_t +\sigma _{xz}\sin\theta_t,\\
\Sigma_z(s)=-(s+\sigma _{xx})\sin\theta_t-\sigma _{zx}\cos\theta_t,\\
\Gamma_{\pm}=\Gamma+2d\beta\text{Im}(f_{\pm}),\label{eq:Gamma}\\
\Delta_{\pm}=\Delta+d\beta\text{Re}(f_{\pm}),\label{eq:Delta} \\
\Gamma^{\prime}=\Gamma+2d\beta\text{Im}(f_{0,yy}),\label{eq:Gamma_s}\\
\Delta^{\prime}=\Delta+d\beta\text{Re}(f_{0,yy}) \label{eq:Delta_s}
\end{eqnarray}
with $\theta_t$ being the angle which the wave transmitted into the slab makes with the normal to the slab boundary and $f_{\pm}$ being the eigenvalues of the matrix $f_{0,ij}(k_0,\omega_0)$ ($i,j=x,z$).\\
The evolution of the MNPs polarization is found as the inverse Laplace transform of Eqs. (\ref{eq:Pix}) - (\ref{eq:Piy}). It is determined by the envelope function of the exciting field, $E_0(t)$. We assume that the system is excited by a rectangular pulse of duration $\tau$, so that
\begin{equation}
E_0(t)=\left\{\begin{array}{cc}
0 &\quad \text{for}\quad t<0, \quad t>\tau,\\
E_0 &\quad \text{for}\quad 0\leq t \leq \tau.
\end{array}\right.
\end{equation}
Then one obtains 
\begin{equation}
P_{0i}^{\alpha}(t)=i\beta E_0 Q_i^{\alpha}(t)
\end{equation}
with
\begin{equation}
Q_i^{\alpha}(t)=R_i^{\alpha}(t)-R_i^{\alpha}(t-\tau)\theta(t-\tau),
\end{equation}
where the superscript $\alpha$ denotes either TM ($p$) or TE ($s$) polarization, $\theta(t)$ is the unit step function, $i=x,z$ for TM polarization and $i=y$ for TE polarization. The functions $R_i^{\alpha}(t)$ are defined as follows
\begin{eqnarray}
R_i^{TM}(t)=\frac{\Sigma_i(0)}{s_-s_+}\nonumber\\
+\frac{\Sigma_i(s_-)}{s_-(s_--s_+)}\exp(s_-t)-\frac{\Sigma_i(s_+)}{s_+(s_--s_+)}\exp(s_+t),\nonumber\\
\end{eqnarray}
\begin{equation}
R_y^{TE}(t)=\frac{1}{s^{\prime}}\left[\exp(s^{\prime}t)-1\right]
\end{equation}
with $s_{\pm}=-(\Gamma_{\pm}/2)+i\Delta_{\pm}$ and $s^{\prime}=-(\Gamma^{\prime}/2)+i\Delta^{\prime}$.\\
\section{Eigenvalues of the field susceptibility tensor}\label{sec:eigenvalues}
The quantity $\beta$ in Eqs. (\ref{eq:Gamma}) - (\ref{eq:Delta_s}) is proportional to the MNP number density or volume fraction that signifies a cooperative effect of MNPs in both the polarization relaxation constant and frequency shift. This effect is determined by the eigenvalues of the field susceptibility Fourier transform tensor which we investigate in this Section.\\
The quantity $\bar{\bf F}({\bf r},{\bf r}^{\prime};\omega)$ can be split into two contributions as
\begin{equation}\label{eq:decompose}
\bar{\bf F}({\bf r},{\bf r}^{\prime};\omega)=\bar{\bf F}^0({\bf r},{\bf r}^{\prime};\omega)+\bar{\bf F}^R({\bf r},{\bf r}^{\prime};\omega),
\end{equation}
where $\bar{\bf F}^0({\bf r},{\bf r}^{\prime};\omega)$ originates from the dipole field of MNPs in free space and $\bar{\bf F}^R({\bf r},{\bf r}^{\prime};\omega)$ results from the MNPs dipole field reflected from the boundaries. Both tensors can be written in the form of the Fourier integrals similar to Eq. (\ref{eq:fourier}), where \cite{Sipe79,Nha96}
\begin{eqnarray}\label{eq:f0}
\bar{\bf f}^0(z,z^{\prime};\kappa_x,\kappa_y;\omega)\nonumber\\
=2\pi i\frac{\tilde{\omega}^2}{W_h}\left[(\hat{s}\hat{s}+\hat{p}_{+}\hat{p}_{+})\theta(Z)e^{iW_hZ}\right.\nonumber\\
+\left.(\hat{s}\hat{s}+\hat{p}_{-}\hat{p}_{-})\theta(-Z)e^{-iW_hZ}\right]\nonumber\\
-4\pi\hat{z}\hat{z}\delta(Z)
\end{eqnarray}
and
\begin{eqnarray}\label{eq:fR}
\bar{\bf f}^R(z,z^{\prime};\kappa_x,\kappa_y;\omega)\nonumber\\
=2\pi i\frac{\tilde{\omega}^2}{W_h}\left[\hat{s}\hat{s}\left(C^-_{ss}(z^{\prime})e^{-iW_hZ}+C^+_{ss}(z^{\prime})e^{iW_hZ}\right)\right.\nonumber\\
+\left.\hat{p}_{+}\hat{p}_{-}C_{p+p-}(z^{\prime})e^{iW_hZ}
+\hat{p}_{-}\hat{p}_{+}C_{p-p+}(z^{\prime})e^{-iW_hZ}\right.\nonumber\\
+\left.\hat{p}_{+}\hat{p}_{+}C_{p+p+}e^{iW_hZ}+\hat{p}_{-}\hat{p}_{-}C_{p-p-}e^{-iW_hZ}\right].\nonumber\\
\end{eqnarray}
Here $Z=z-z^{\prime}$, $\tilde{\omega}=\omega/c=2\pi/\lambda$ with $c$ being the speed of light in vacuum, $W_h=(\tilde{\omega}^2\epsilon_h-\kappa^2)^{1/2}$ is the $z$-component of the plane wave in the host material, $\kappa^2=\kappa_x^2+\kappa_y^2$ is the squared modulus of the wave vector component parallel to the slab surfaces, $\bm{\kappa}$, the unit vectors $\hat{z}$, $\hat{\kappa}$ and $\hat{s}=\hat{\kappa}\times \hat{z}$ are oriented along the corresponding directions and $\theta(Z)$ is the unit step function with $\theta(0)=1/2$ that ensures the symmetry of the dipole field in free space. The other notations are as follows
\begin{equation}
\hat{p}_{\pm}=(\tilde{\omega}\sqrt{\epsilon_h})^{-1}(\kappa\hat{z}\mp W_h\hat{\kappa}),
\end{equation}
\begin{equation}
C_{ss}^-(z)=\frac{r_1^se^{iW_h(d-2z)}+r_1^sr_2^se^{2iW_hd}}{1-r_1^sr_2^se^{2iW_hd}},
\end{equation}
\begin{equation}
C_{ss}^+(z)=\frac{r_1^se^{iW_h(d+2z)}+r_1^sr_2^se^{2iW_hd}}{1-r_1^sr_2^se^{2iW_hd}},
\end{equation}
\begin{equation}
C_{p+p-}(z)=\frac{r_1^pe^{iW_h(d+2z)}}{1-r_1^pr_2^pe^{2iW_hd}},
\end{equation}
\begin{equation}
C_{p-p+}(z)=\frac{r_2^pe^{iW_h(d-2z)}}{1-r_1^pr_2^pe^{2iW_hd}},
\end{equation}
and
\begin{equation}
C_{p-p-}=C_{p+p+}=\frac{r_1^pr_2^pe^{2iW_hd}}{1-r_1^pr_2^pe^{2iW_hd}},
\end{equation}
where
\begin{eqnarray}
r_1^s=\frac{W_h-W_1}{W_h+W_1},\\
r_1^p=\frac{\epsilon_1 W_h-\epsilon_hW_1}{\epsilon_1 W_h+\epsilon_hW_1},\\
r_2^s=\frac{W_h-W_2}{W_h+W_2},\\
r_2^p=\frac{\epsilon_2 W_h-\epsilon_hW_2}{\epsilon_2 W_h+\epsilon_hW_2},
\end{eqnarray}
are the Fresnel reflection coefficients for $s$ and $p$ polarizations at the medium 1/slab and slab/medium 2 interfaces, respectively, in the approximation of small MNP volume fraction with $W_1=(\tilde{\omega}^2\epsilon_1-\kappa^2)^{1/2}$ and $W_2=(\tilde{\omega}^2\epsilon_2-\kappa^2)^{1/2}$. In Eqs. (\ref{eq:f0}) and (\ref{eq:fR}), the parts of the tensors proportional to $\hat{s}\hat{s}$ describe the electric field for TE polarization ($s$ polarization) of the dipoles, whereas the other parts proportional to $\hat{p}_{\pm}\hat{p}_{\pm}$ are relevant to TM polarization ($p$ polarization) of the dipoles.\\
In the approximation under consideration one can take the field susceptibility tensor in the limit $d\rightarrow 0$. This also implies $z\rightarrow z^{\prime}\rightarrow 0$ provided $z\neq z^{\prime}$ that excludes self-action of MNPs. In the case under discussion $\hat{\kappa}=\hat{x}$ and $\hat{s}=\hat{y}$ that allows one to write $f_{0,ij}\equiv 2\pi i\phi_{ij}$ with
\begin{eqnarray}
\phi_{xx}=\frac{W_h}{\epsilon_h}\frac{(1-r_1^p)(1-r_2^p)}{1-r_1^pr_2^p},\label{eq:fxx}\\
\phi_{zz}=\frac{k_0^2}{\epsilon_hW_h}\frac{(1+r_1^p)(1+r_2^p)}{1-r_1^pr_2^p},\label{eq:fzz}\\
\phi_{xz}=-\phi_{zx}=\frac{k_0}{\epsilon_h}\frac{r_2^p-r_1^p}{1-r_1^pr_2^p}\label{eq:fxz},
\end{eqnarray}
\begin{equation}\label{eq:fyy}
\phi_{yy}=\frac{\tilde{\omega}_0^2}{W_h}\frac{(1+r_1^s)(1+r_2^s)}{1-r_1^sr_2^s}=\frac{2\tilde{\omega}_0^2}{W_1+W_2}
\end{equation}
and the other matrix elements $\phi_{ij}$ are equal to zero.\\
The eigenvalues of the field susceptibility tensor are found as $f_{\pm}=2\pi i \phi_{\pm}$ and $f_{0,yy}=2\pi i\phi_{yy}$, where $\phi_{\pm}$ satisfy the equation
\begin{equation}\label{eq:quadratic}
\phi_{\alpha}^2-(\phi_{xx}+\phi_{zz})\phi_{\alpha}+\phi_{xx}\phi_{zz}+\phi_{xz}^2=0,\quad \alpha=\pm.
\end{equation}
It follows from here that 
\begin{eqnarray}\label{eq:sigma}
\phi_-+\phi_+=\phi_{xx}+\phi_{zz}\nonumber\\
=\frac{2\epsilon_1\epsilon_2}{\epsilon_1W_2+\epsilon_2W_1}\left(\frac{W_1W_2}{\epsilon_1\epsilon_2}+\frac{k_0^2}{\epsilon_h^2}\right)
\equiv\sigma_1+i\sigma_2
\end{eqnarray}
and 
\begin{equation}\label{eq:product}
\phi_-\phi_+=\phi_{xx}\phi_{zz}+\phi_{xz}^2=\frac{k_0^2}{\epsilon_h^2},
\end{equation}
where $\sigma_1$ and $\sigma_2$ are the real and imaginary parts of the right-hand side part of Eq. (\ref{eq:sigma}), respectively. The latter equation allows one to write
\begin{equation}
\phi_-=p\exp(i\psi)
\end{equation}
and
\begin{equation}
\phi_+=q\exp(-i\psi),
\end{equation}
where $p$ and $q$ are real quantities such that $pq=k_0^2/\epsilon_h^2$ and $\psi$ is a real phase. Then, separating the real and imaginary parts in Eq. (\ref{eq:sigma}), one obtains
\begin{equation}
(p+q)\cos\psi=\sigma_1
\end{equation}
and
\begin{equation}
(p-q)\sin\psi=\sigma_2.\\
\end{equation}
In what follows, we assume that the superstrate material is dielectric, while the substrate can be either a dielectric or a metal. In such a case both $\epsilon_1$ and $W_1$ are real and positive quantities.\\
When the substrate is a dielectric both $\epsilon_2$ and $W_2$ are real and $\sigma_2=0$, so that either $\psi=0$, or $p=q$. Correspondingly, the roots $\phi_-$ and $\phi_+$, which determine the cooperative effect in TM polarization, are either real or complex conjugate to each other, depending on the sign of the discriminant of Eq. (\ref{eq:quadratic}). In the first case the cooperative effect takes place for the relaxation rate, Eq. (\ref{eq:Gamma}), only. In the latter case it contributes to both the relaxation rate and the frequency shift. The magnitude of the effect is reduced as compared to the previous case.\\
When the substrate is a metal, $\epsilon_2<0$ in neglect of its imaginary part, that is a good approximation in the optical spectral range, and $W_2$ is imaginary. In this case, $\phi_-$ and $\phi_+$ are complex quantities and contribute to both the relaxation rate and the frequency shift. A similar analysis can be performed for the quantity $\phi_{yy}$, Eq. (\ref{eq:fyy}), which determines the cooperative effect of MNPs in TE polarization.\\
There can be, however, a particular case where the incident field is excited in the attenuated total reflection (ATR) configuration by TM polarized light. It can be realized, for example, when the incident light passes through a prism with the refractive index $n_p>\sqrt{\epsilon_1}$ suspended above the metal substrate at a distance comparable with the wavelength (Otto configuration) \cite{Raether88}. Then $W_1$ is imaginary as well and a resonance in the quantity $\phi_-+\phi_+$, Eq. (\ref{eq:sigma}), specified by the condition
\begin{equation}
\text{Im}(\epsilon_1W_2+\epsilon_2W_1)=0,
\end{equation}
becomes possible which corresponds to the excitation of surface plasmon polaritons (SPPs) at the interface between the substrate and superstrate \cite{Raether88}. At exact resonance the quantity $\epsilon_1W_2+\epsilon_2W_1$ is real and determined by the imaginary part of $\epsilon_2$. In such a case, one of the roots of Eq. (\ref{eq:quadratic}) is much larger in absolute value than the other one, as it follows from Eq. (\ref{eq:product}), and one can expect an enhanced cooperative effect of MNPs.\\
\section{Radiation field in the slab}\label{sec:radiation}
The electric field of the cooperative radiation emitted by MNPs in the slab is found as [see Eq. (\ref{eq:integral})]
\begin{equation}\label{eq:integral_T}
{\bf E}_r(z,t)=\int_{-d/2}^{d/2} \bar{\bf f}(z,z^{\prime};k_0,0;\omega_0){\bf P}_0(z^{\prime},t)dz^{\prime}.
\end{equation}
Taking into account that the nanocomposite is optically thin as before, one finds 
\begin{equation}
{\bf E}_{ri}^{\alpha}(t)=-2\pi \beta dE_0\sum_j\phi_{ij}Q_j^{\alpha}(t).
\end{equation}
and the relative intensity of the radiated field
\begin{equation}\label{eq:intensity}
\frac{I_r^{\alpha}(t)}{I_0}=\frac{\mid E_r^{\alpha}(t)\mid ^2}{\mid E_0\mid ^2}=4\pi^2\beta^2d^2\sum_i\left |  \sum_{j}\phi_{ij}Q_j^{\alpha}(t)\right| ^2.\\
\end{equation}
Equation (\ref{eq:intensity}) can be represented in terms of the dimensionless parameters
\begin{eqnarray}
\tilde{\Gamma}_{\pm}=\Gamma_{\pm}/\Gamma=1+2\eta\text{Re}(\tilde{\phi}_{\pm}),\label{eq:gammapm}\\
\tilde{\Delta}_{\pm}=\Delta_{\pm}/\Gamma=\tilde{\Delta}-\eta\text{Im}(\tilde{\phi}_{\pm}),\\
\tilde{\Gamma}^{\prime}=\Gamma^{\prime}/\Gamma=1+2\eta\text{Re}(\tilde{\phi}_{yy}),\\
\tilde{\Delta}^{\prime}=\Delta^{\prime}/\Gamma=\tilde{\Delta}-\eta\text{Im}(\tilde{\phi}_{yy})\label{eq:deltap}
\end{eqnarray}
with $\tilde{\Delta}=\Delta/\Gamma$, $\tilde{\phi}_{\nu}=\phi_{\nu}/\tilde{\omega}_0$ and the dimensionless function $\tilde{Q}(t)$ as follows
\begin{equation}\label{eq:intensity_til}
\frac{I_r^{\alpha}(t)}{I_0}=\eta^2\sum_i\left |  \sum_{j}\tilde{\phi}_{ij}\tilde{Q}_j^{\alpha}(t)\right| ^2
\end{equation}
that reveals a governing role of the parameter $\eta=2\pi\beta d\tilde{\omega}_0/\Gamma$ in the cooperative radiation.\\
Let us note that in the approximation under consideration the relaxation rates and frequency shifts, Eqs. (\ref{eq:gammapm}) - (\ref{eq:deltap}), linearly depend on the parameter $\eta\sim\beta d\sim fd\sim Nd$, i.e. on the number of MNPs per unit area of the slab. On the other hand, the radiated intensity, Eq. (\ref{eq:intensity_til}), is proportional to this number squared that is a feature of the Dicke superradiance.\\
\section{Numerical results}\label{sec:numerical}
We illustrate the theory developed above by the numerical calculations for silver nanoparticles distributed throughout a host material with $\epsilon_h=1.5^2$ in contact with vacuum as a superstrate ($\epsilon_1=1$). We adopt the Drude model, Eq. (\ref{eq:drude}), for the dielectric function of MNPs with $\epsilon_{\infty}=5$, $\omega_p=14.0\times 10^{15}$ s$^{-1}$ and $\Gamma=0.32\times 10^{14}$ s$^{-1}$ \cite{Shalaev10}. The corresponding wavelength of the LSP resonance is $\lambda_0=2\pi c/\omega_0\approx 415$ nm. When the substrate is a metal one can excite SPPs by TM polarized light passing through a prism at the incidence angle $\theta_i$ satisfying the equation \cite{Raether88}
\begin{equation}
n_p\sin\theta_i=\sqrt{\frac{\epsilon_2}{\epsilon_2+1}},
\end{equation}
where $n_p$ is the prism refractive index. For a silver substrate and $n_p=1.5$ this gives $\theta_i\approx 49^{\circ}$ for $\lambda=415$ nm.\\
\begin{figure}
\includegraphics[width=\linewidth]{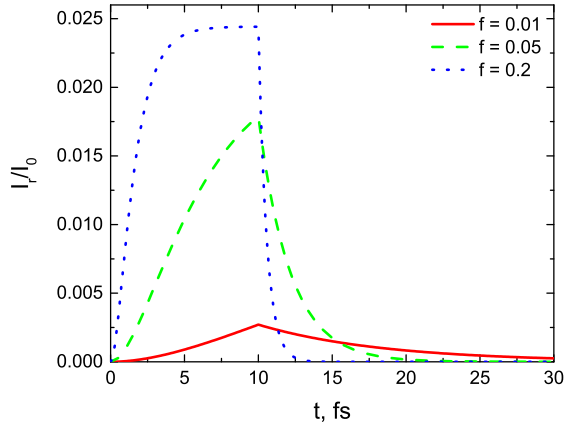}
\caption{\label{fig:TE} The Dicke superradiance in TE polarization from a nanocomposite with Ag nanoparticles as a function of time for different volume fractions of MNPs. $\epsilon_1=1$, $\epsilon_2=1.7^2$, $\epsilon_h=1.5^2$, $d=50$ nm, $\lambda=415$ nm, $\theta_i=10^{\circ}$ and $\tau=10$ fs.}
\end{figure} 
\begin{figure}
\includegraphics[width=\linewidth]{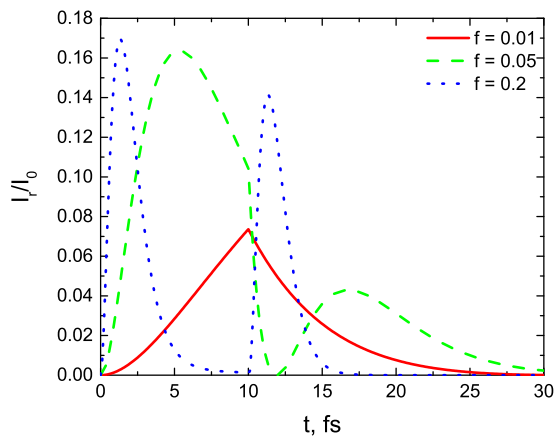}
\caption{\label{fig:TM} Same as in Fig. \ref{fig:TE}, but for TM polarization.}
\end{figure} 
\begin{figure}
\includegraphics[width=\linewidth]{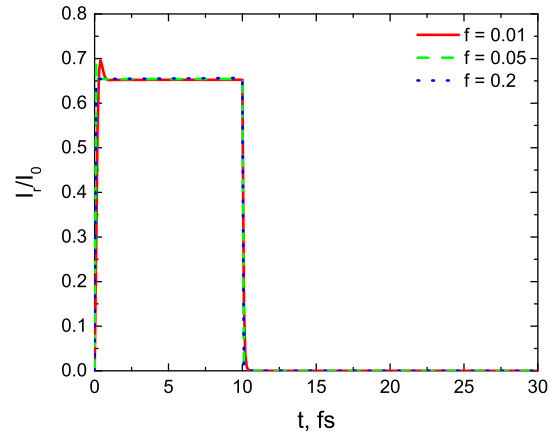}
\caption{\label{fig:SPP} The Dicke superradiance excited in TM polarization in the ATR configuration for different volume fractions of MNPs. $n_p=1.5$, $\theta_i=49^{\circ}$; the other parameters are the same as in Fig. \ref{fig:TE}.}
\end{figure} 
Figure \ref{fig:TE} shows the evolution of the cooperative radiation intensity excited by TE polarized light. The calculations are carried out for a dielectric substrate with $\epsilon_2=1.7^2$ that is close to that of Al$_2$O$_3$. One can see that both the signal growth rate and its intensity gradually increase with the volume fraction of MNPs during the pumping pulse. When the external field is switched off the same tendency is observed for the signal decay.\\
The dynamics of the superradiance demonstrated in Fig. \ref{fig:TM} for TM polarization is qualitatively different. For a relatively small volume fraction of MNPs the overall behavior is similar to that for TE polarization. However for larger values of $f$ there develops a peak within the action of the pumping pulse. There is also a replica of this peak, associated with the termination of the pulse, which has a smaller amplitude. The intensities of the peaks are substantially larger than the signal intensity in TE polarization.\\
Figure \ref{fig:SPP} shows the radiation intensity evolution for a nanocomposite slab deposited onto an Ag substrate. The calculations are carried out for the excitation of SPPs in the ATR configuration at the interface between the substrate and superstrate. In this scenario, the cooperative response of MNPs is so fast that the signal almost repeats the shape of the pumping pulse even for small $f$. Its intensity is enhanced in comparison with the previous two cases.\\
\section{Conclusion}\label{sec:conclusion}
In this paper, we have predicted the Dicke superradiance which can be observed from an optically thin plasmonic nanocomposite slab. We have developed the theory of this effect basing on the Green's function approach which properly accounts the local field in the nanocomposite. \\
Excitation of the system in both TE and TM polarizations as well as in the ATR configuration have been considered. The superradiance intensity is shown to be proportional to the square of the number of MNPs per unit area of the slab, while the relaxation times and the resonance detunings linearly depend on this quantity. It is demonstrated that the relaxation rates are significantly enhanced when SPPs are excited at the boundary between the substrate and superstrate in the ATR configuration.\\
The research reported in this paper reveals that the phenomenon of superradiance can be observed in the femtosecond time domain due to extremely fast relaxation processes in MNPs. These findings can form a basis for novel ultra-fast optical techniques which exploit the Dicke superradiance. In particular, they can be used to engineer femtosecond pulses of coherent radiation using a rather simple approach. \\
\section*{Acknowledgments}
This research is kindly supported by the IE-Industrial Elektronik project (SFD-17-0036) which has received EU co-financing from the European Social Fund.

\end{document}